\documentclass[prl,twocolumn,showpacs,preprintnumbers,amsmath,amssymb]{revtex4}
\usepackage{dcolumn}
\usepackage{bm}
\usepackage{epsfig}

\begin{document}

\preprint{}

\title{Severe Fermi Surface Reconstruction at a Metamagnetic-Transition in Ca$_{2-x}$Sr$_x$RuO$_4$ \\(for $0.2 \leq x \leq 0.5$)}

\author{L. Balicas,$^1$ S. Nakatsuji,$^2$ D. Hall,$^{3}$ H. Lee,$^{4}$ Z. Fisk$^{4}$, Y. Maeno$^2$, and D. J. Singh$^{5}$}
\affiliation{$^1$National High Magnetic Field Laboratory, Florida
State University, Tallahassee-FL 32306, USA}
\affiliation{$^2$Department of Physics, Graduate School of Science
Kyoto University, Kyoto 606-8502, Japan} \affiliation{$^3$
American Physical Society, 1 Research Road, P. O. Box 9000, Ridge,
NY, 11961.} \affiliation{$^4$Department of Physics, University of
California, Davis, California, 95916}
\affiliation{$^5$Condensed Matter Sciences Division, Oak Ridge
National Laboratory, Oak Ridge, TN 37831-6032}

\date{\today}%
\begin{abstract}
We report an electrical transport study in
Ca$_{2-x}$Sr$_{x}$RuO$_4$ single crystals at high magnetic fields
($B$). For $x =0.2$, the Hall constant $R_{xy}$ decreases sharply
at an anisotropic metamagnetic (MM) transition reaching its value
for Sr$_2$RuO$_4$ at high fields. A sharp decrease in the $A$
coefficient of the resistivity $T^2$-term and a change in the
structure of the angular magnetoresistance oscillations (AMRO) for
$B$ rotating in the planes, confirms the reconstruction of the
Fermi surface (FS). Our observations and LDA calculations indicate
a strong dependence of the FS on the Ca concentration and suggest
the coexistence of itinerant and localized electronic states in
single layered ruthenates.

\end{abstract}

 \maketitle

The single layered ruthenates Ca$_{2-x}$Sr$_x$RuO$_4$ have a quite
complex phase-diagram as function of doping $x$
\cite{satoru,satoruprb}. While Sr$_2$RuO$_4$ is a rare example of
a well-defined two-dimensional Fermi liquid (FL) \cite{yoshi,andy}
displaying spin-triplet superconductivity, the complete
replacement of Sr by the isovalent element Ca produces to the Mott
insulator Ca$_2$RuO$_4$ \cite{satorujpsj,satoru}. A series of correlated
metallic states are protruded between both extremes.

In this system the relevant $4d$-orbitals belong to the
$t_{2g}$-subshell and are degenerate. The planar structure conducts
to very weak hybridization between orbitals which have either even
($d_{xy}$) or odd ($d_{yz}, d_{zx}$) parity under the reflection
$z \rightarrow -z$. These orbitals lead in Sr$_2$RuO$_4$ to a
Fermi surface composed by three warped cylinders \cite{bergemann}.
The $\alpha$ and $\beta$ sheets arise from 1D chains of
$d_{yz};d_{zx}$ orbitals while the $\gamma$ FS sheet originates
from a 2D network of planar $d_{xy}$ orbitals. Starting from the
undistorted Sr$_2$RuO$_4$, Ca substitution initially stabilizes
the rotation of RuO$_6$ octahedra. This induces an enhancement of
both the low $T$ magnetic susceptibility $\chi(T)$ and the
Sommerfeld coefficient characterizing the electronic contribution
to the specific heat \cite{satoru2}. Both quantities display a
pronounced maximum at a critical value $x_c = 0.5$ where the
crystallographic structure of the Ca$_{2-x}$Sr$_{x}$RuO$_{4}$
series changes from tetragonal to orthorhombic through a
second-order structural transition \cite{satoru}. For stronger Ca
concentrations and as we approach the Mott transition at $x
\lesssim 0.2$, $\chi(T)$ is strongly suppressed by the emergence
of AF correlations.

This complex evolution has led to various theoretical proposals,
in particular to the idea that some of the $d$-orbitals display
localized spin and orbital degrees of freedom, while others
provide itinerant carriers. This so-called orbital-selective Mott
transition (OSMT) \cite{anisimov} could, for instance, explain the
experimental observation of an effective spin $S$ very close to
the value 1/2 for $0.2\leq x \leq 0.5$ \cite{satoru2} which is
difficult to understand within an entirely itinerant description.
However the concept of orbital-selective Mott-transition was
challenged by the the dynamical mean field (DMFT) calculations of
Liebsch \cite{liebsch} which suggest that two bands of different
widths coupled by electron-electron interactions always undergo a
single simultaneous Mott transition. Koga \emph{et al}. claim that
this apparent discrepancy can be solved by taking into account a
finite Hund coupling in Hubbard models considering multiple bands
with different bandwidths \cite{koga}. In contrast, according to
Okamoto and Millis \cite{okamoto} Hund coupling would in fact
stabilize a given system against orbital ordering preventing the
occurrence of the OSMT. Finally, a recent photoemission study in
Ca$_{1.5}$Sr$_{0.5}$RuO$_4$ finds that the geometry of the
$\alpha$ and $\beta$ FS sheets remains almost unchanged respect to
Sr$_2$RuO$_4$, with the $\gamma$ sheet exhibiting a hole-like FS
in contrast to being electron-like in the pure Sr compound
\cite{arpes}. The observation of \emph{all} three volume
conserving FS sheets would demonstrate the absence of OSMT in
Ca$_{1.5}$Sr$_{0.5}$RuO$_4$.

To clarify the possible existence of localized carriers within a
metallic $d$-electron system as proposed by the orbital-selective
Mott-transition model, we performed a detailed electrical
transport study at high magnetic fields in Ca$_{2-x}$Sr$_x$RuO$_4$
single crystals for $x = 0.2$ and 0.5. In particular, we studied
the Hall effect and the angular dependence of the
magnetoresistance or AMRO, which is a technique commonly used to
define the relatively simple two-dimensional (2D) FS of layered
organic metals \cite{wosnitzaishiguro}, and which has already
provided valuable information about the FS of Sr$_2$RuO$_4$
\cite{ohmichiamro, ohmichiinplane}. Our measurements reveal a
profound modification of the original FS of Sr$_2$RuO$_4$ upon Ca
doping and a severe reconstruction of the FS of $x = 0.2$ at a
metamagnetic transition. These observations are consistent with
the proposed existence of localized $d$-electronic states in
Ca$_{2-x}$Sr$_x$RuO$_4$ (for $0.2 \leq x \leq 0.5$).

\begin{figure}[t]
\begin{center}
\epsfig{file=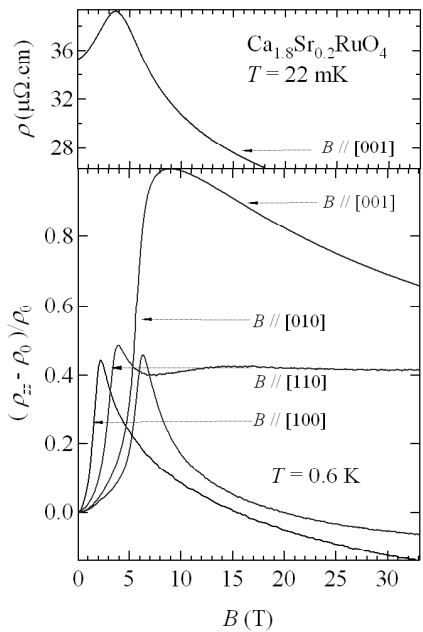, width=5.5 cm} \caption{Upper
panel: In-plane resistivity $\rho$ for a Ca$_{1.8}$Sr$_{0.2}$
RuO$_4$ single crystal as a function of the field $B$ applied
along the inter-plane direction at $T=22$ mK. Lower panel:
Inter-plane resistivity $\rho_{zz}$ normalized respect to its zero
field value $\rho_{0}$, as a function of the external field $B$
oriented along different crystallographic orientations and at $T
\simeq 0.6$. In both panels a pronounced peak is observed at an
anisotropic metamagnetic transition.}
\end{center}
\end{figure}
\begin{figure}[t]
\begin{center}
\epsfig{file=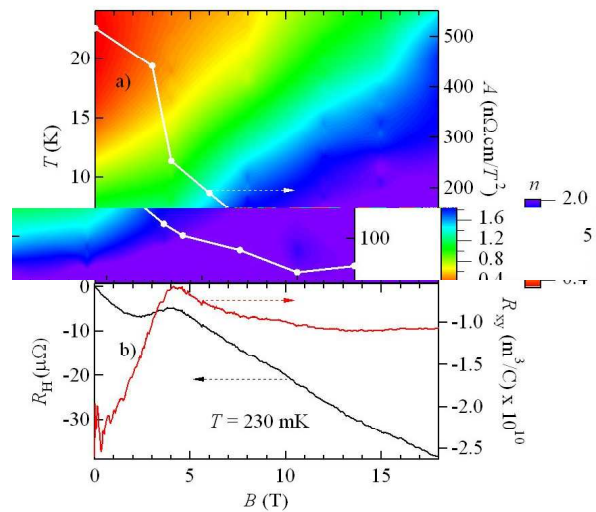, width=7.5 cm} \caption{a) The
 exponent $n = \partial \ln(\rho-\rho_0)/ \partial(T) $ of the in-plane
 resistivity $\rho$ in the $B-T$ plane. The Fermi-liquid ground state is always observed at lower $T$s.
 The white symbols describe the evolution of the $A (= (\rho-\rho_0)/T^2)$ coefficient as a function $B$.
 b) The Hall resistance $R_H$ and the Hall constant $R_{xy}$ as a function of $B$ at $T = 230$ mK. Notice the sharp decrease in $R_{xy}$ at the metamagnetic transition. }
\end{center}
\end{figure}

The upper panel of Fig. 1 shows the in-plane resistivity of Ca$_{1.8}$Sr$_{0.2}$RuO$_{4}$
single crystal at $T = 22$ mK and as a function of the field $B$ applied along the inter-plane direction.
While the lower panel of Fig. 1 shows the relative change of its inter-plane resistivity
($\rho_{zz}$ - $\rho_0$)/$\rho_0$ (where $\rho_0$ is the residual
resistivity at $B=0$)  as a function of $B$ at $T = 600$ mK
and for four orientations respect to $B$ (as indicated in the figure).
For all orientations both $\rho$ and  $\rho_{zz}$
display pronounced positive magnetoresistivity preceding a
sharp peak beyond which the system displays a remarkable negative
magnetoresistance (the exact position in $B$ is sample dependent).
Notice how $\rho$ at high fields decreases beyond its value at $B=0$.
This peak is produced by a metamagnetic (MM) transition, i.e., a rapid increase in magnetization
between two distinct paramagnetic states, the first one
characterized by short-range antiferromagnetic correlations and the
second one dominated by ferromagnetic-like interactions \cite{satoru2}. Both, the position of
the metamagnetic critical field $B_{\text{MM}}$ (as defined by the
peak in the resistivity) as well as the background
magnetoresistivity display a remarkable orientation dependence.

\begin{figure}[t]
\begin{center}
\epsfig{file=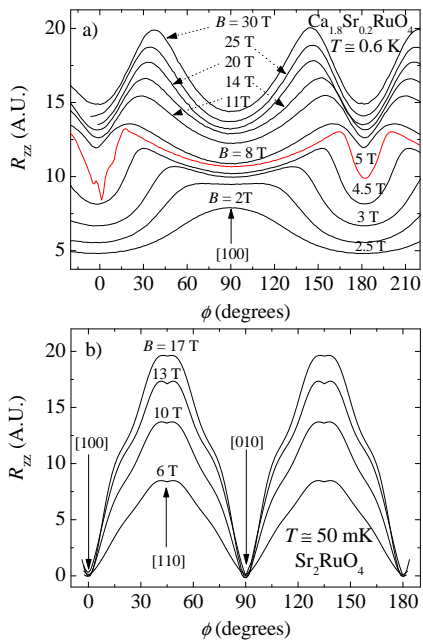, width=5.5 cm} \caption{ a)
Inter-plane resistance $R_{zz}$ as a function of the angle $\phi$
between $B$ and an in-plane axis for a
Ca$_{1.8}$Sr$_{0.2}$RuO$_{4}$ single crystal at $T=0.6$ K. Notice
how the low field two-fold periodicity progressively becomes
four-fold at high fields after the metamagnetic transition is
crossed. b) $R_{zz}$ for a Sr$_{2}$RuO$_{4}$ single crystal as
function of the angle $\phi$ and for several values of $B$ and at
$T\simeq 50$ mK. Notice the progressive emergence of a new
periodicity at higher fields. In both graphs, all curves are
vertically displaced for clarity.}
\end{center}
\end{figure}

The color plot in Fig. 2(a) shows the exponent $n = \partial
\ln(\rho-\rho_0)/ \partial(T) $ of the in-plane resistivity $\rho$
in a limited range of the $T-B$ space. The purple regions emerging
at low temperatures reveal the existence of a Fermi liquid (FL)
ground state ($n=2$). Notice how the characteristic temperature
$T_{\text{FL}}$ below which FL-like behavior is observed increases
with $B$. The white line and markers indicate the evolution in $B$
of the resistivity's $A = (\rho-\rho_0)/T^2$ coefficient which
shows a pronounced decrement at the transition. At $B=0$, $A$ is
proportional to the square of the density of states at the Fermi
level $\rho(\epsilon_F)^2$ via the Kadowacki-Woods ratio. This
reduction in $\rho(\epsilon_F)$, confirmed by heat capacity
measurements \cite{bradenhc}, points towards a reconfiguration of
the FS at the MM-transition. Furthermore, the decrease in $A$
coupled to an increase in $T_{\text{FL}}$ has usually been taken
as an indication of the close proximity of a given system to a
quantum critical point at $B = 0$ \cite{taillefer}, which in our
case is probably associated with the Mott transition. The fact
that $A$ and $T_{\text{FL}}$ as well as the lattice constants
\cite{braden} change continuously with $B$, while $\rho_{zz}$
shows a discontinuity at the  metamagnetic transition suggests
that it corresponds to a percolation-like first-order transition
as proposed by Ref. \cite{sigrist}. Fig. 2(a) shows no evidence of
critical behavior as expected for a first-order metamagnetic
transition. While Fig. 2(b) shows the most relevant result of this
study, the Hall resistance $R_H = V_H/I$, where $V_H$ is the Hall
voltage and $I$ the electrical current, and the Hall constant
$R_{xy}=R_H \cdot t / B$, where $t$ is the sample thickness, as a
function of $B$ at $T = 230$ mK. At such low $T$s, Ref.
\cite{andyHall} has demonstrated that one reaches the
elastic-scattering regime where $R_H$ is determined mainly by the
FS topography rather than by the details of the scattering
mechanism. Notice how the absolute value of $R_{xy}$ decreases
through the metamagnetic transition from nearly $2.5 \pm 0.5
\times 10^{-10}$ m$^3$/C at very low fields and saturates at the
value of $\simeq$ $1 \pm 0.05 \times 10^{-10}$ m$^3$/C, which is
precisely the value reported for Sr$_2$RuO$_4$ \cite{andyHall}.
This indicates an important modification of the original band
structure of Sr$_2$RuO$_4$ upon Ca doping as well as a severe
reconstruction of the FS at the metamagnetic transition. The Hall
mobility $R_{xy}/ \rho$ decreases rapidly from $\sim 6.5$
cm$^2$/Vs at $B = 0$ to a value of about $\sim 1.5$ cm$^2$/Vs at
$B_{\text{MM}}$ saturating at high $B$s to a value of $\sim 4.0$
cm$^2$/Vs reflecting an increase in scattering events. Thus, the
increase in conductivity above the transition cannot be attributed
to an increase in mobility, but it reflects perhaps the reduction
in the effective mass of the charge carriers. Notice that in
multi-band systems such as the ruthenates, whose FS may be
composed by both electron and hole FS sheets, $R_{xy}$ is no
longer $\propto n_c^{-1}$ where $n_c$ is the number of charge
carriers. Although the abrupt change in $R_{xy}$ cannot be
interpreted as a change in FS volume, this metamagnetic behavior
bears similarities with the metamagnetism followed by colossal
magnetoresistive-like behavior seen in the Ca$_3$Ru$_2$O$_7$
system and which is ascribed to the destabilization of an
orbital-ordered state \cite{gang}. Similarly, in our case it could
also indicate the existence of localized $d$-electron states that
are destabilized by the external field.

\begin{figure}[t]
\begin{center}
\epsfig{file=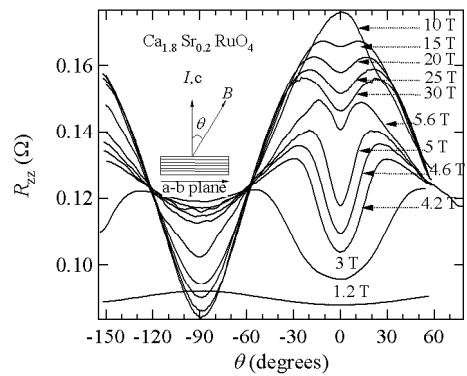, width=6 cm} \caption{Inter-plane
resistance $R_{zz}$ as a function of the angle $\theta$ between
the external magnetic field $B$ and the inter-plane c-axis for a
Ca$_{1.8}$Sr$_{0.2}$RuO$_4$ single crystal at $T = 0.6$ K.}
\end{center}
\end{figure}

Figure 3(a) shows the inter-plane resistance $R_{zz}$ of a
Ca$_{1.8}$Sr$_{0.2}$RuO$_4$ single crystal at $T = 0.6$ K and as a
function of the azimuthal angle $\phi$ between $B$ and an in-plane
axis for several values of $B$ below and above $B_{\text{MM}}$(as
indicated in the figure). All curves are vertically displaced for
clarity. At low fields, $R_{zz}(\phi)$ is essentially two-fold. A
similar magnetoresistive effect, but displaying four-fold
periodicity (that reflects the angular symmetry of the CuO$_2$
planes which leads to a four-fold modulation in $k_F$, $v_F$ and
$\tau$), was reported in the tetragonal compound
Tl$_2$Ba$_2$CuO$_6$ \cite{nigelprl} as well as in Sr$_2$RuO$_4$
\cite{ohmichiinplane}. Here, the two-fold sinusoidal-like
dependence, which reflects the symmetry of the orthorhombic
structure, progressively disappears as the MM-transition is
approached. As $B$ further increases a new, nearly four-fold,
periodicity emerges. This unique evolution of the AMRO data across
the MM-transition is consistent with the reconstruction of the FS
revealed by $R_{xy}$. It cannot be simply explained in terms
semiclassical orbits exploring the geometry of the FS in greater
detail at higher fields. To illustrate this point in Fig. 3(b) we
show $R_{zz}(\phi)$ for a Sr$_2$RuO$_4$ single crystal at $T = 50$
mK and for several values of $B$ (traces are vertically displaced
for clarity). At lower $B$s one clearly observes the four-fold
AMRO but as $B$ increases its amplitude increases with a new
periodicity emerging and \emph{overlapping} it. The new structure
reflecting the warping of the FS, does \emph{not} emerge in
detriment of the lower field one(s), contrary to the $x=0.2$ case.
More importantly, even above the metamagnetic transition the
symmetry of the FS for $x=0.2$ remains essentially two- and not
four-fold.

A two-dimensional FS composed of three open cylinders, as seen by
ARPES for $x=0.5$, would generate a distinctive polar AMRO
($\rho_{zz} (\theta, B)$ where $\theta$ is the angle between $B$
and the c-axis), as seen for instance, in Sr$_2$RuO$_4$
\cite{ohmichiamro}. For very low levels of disorder, one should
observe a series of peaks, periodic in $\tan(\theta)$, due to the
Yamaji effect \cite{yamaji}. While for disordered systems one
might observe only a small amplitude single sinusoidal component
showing a minimum for $B \parallel$ c-axis. Nevertheless, as seen
in Fig. 4 which shows $R_{zz}(\theta)$ at $T=0.6$ K and for
several field values, a very pronounced AMRO is indeed observed
for $x = 0.2$ but it \emph{does not} display an angular structure
compatible with the above discussion. A quite similar but far less
pronounced AMRO (as well as MM behavior) is observed also in $x =
0.5$ (not shown here). The origin of the observed dependence
remains unclear, although it is similar to that of the so-called
Lebed oscillations seen in quasi-one-dimensional organic
conductors \cite{lebed}. Were this their correct description, it
would imply that the 2D $\gamma$ FS sheet has moved below
$\rho(\epsilon_F)$ while the $\alpha$ and/or the $\beta$ Q1D FS
sheets would survive this level of doping. Quasi-one-dimensional
FS sheets with a certain degree of nesting would explain the
development of AF correlations and would be compatible with the
two-fold azimuthal AMRO. Notice that according to Ref. \cite{fang}
the rotation of the RuO$_6$ octahedra is expected to seriously
reduce the occupation of the $\gamma$ band. In any case, this
unusual polar AMRO  implies that high levels of Ca doping modifies
the original FS of Sr$_2$RuO$_4$ at the point that it no longer
reveals a clear 2D character, in sharp contrast with the ARPES
measurements \cite{arpes}.

In order to understand the evolution of the FS as a function of Ca
doping, LDA calculations were done using the LAPW method and the
neutron crystal structures \cite{braden} of $x=0$ at 180 K (S0
short octahedra, Mott insulator), of $x=0$ at 400 K (L0, long
octahedra, metal) and $x=0.5$ at 10 K (L5). For the S0 an
antiferromagnetic state, $m \sim 2 \mu_B$ is predicted; the
electronic structure has very narrow bands at $E_F$, favorable for
the observed Mott insulating ground state. The L0 structure is an
itinerant ferromagnet, $m \sim 1 \mu_B$ in the LDA. Electric field
gradients (EFG) are sensitive probes of orbital occupation. A
large change is found between the L0 and S0 structures: the
average Ru EFG's are -4.9x10$^{21}$ V/m$^2$ and +4.3x10$^{21}$
V/m$^2$, for paramagnetic L0 and antiferromagnetic S0,
respectively. The L5 structure is weakly ferromagnetic, presumably
due to neglect of quantum critical fluctuations in the LDA.
$x=0.2$ is on the borderline between these two states, i.e. it is
in a long octahedra structure intermediate between L0 and L5, is
on the verge of the long to short crossover. The L0 FS is complex
due to zone folding and lowered symmetry, but, it has two
contributions, heavy bands (H), that do not contribute to
transport and an itinerant component. The H-FS is strongly
affected by magnetic ordering, seen in a reduction by a factor of
more than 2 in the EFG (to -2.3x10$^{21}$ V/m$^2$) when
ferromagnetic, showing a change in orbital population. This is
accompanied by a large change in the density of states:
$N(E_F)$=2.5 eV$^{-1}$ per Ru per spin (paramagnetic) 1.3
eV$^{-1}$ (majority) and 2.9 eV$^{-1}$ (minority), even though the
in-plane transport function $Nv_x^2$, which reflects the itinerant
sheets, is the same to better than 10\% for the paramagnetic,
ferromagnetic majority and ferromagnetic minority FS's. If the
situation is the same at $x$=0.2 as expected, the implication is
that at the metamagnetic transition the heavy part of the FS is
removed for majority carriers, consistent with the experimental
results.

In summary, the Fermi surface of the Ca$_{2-x}$Sr$_x$RuO$_4$
system is strongly dependent on $x$. The anomalous polar AMRO and
the severe field-induced Fermi surface reconstruction observed at
the metamagnetic transition of $x=0.2$, could be evidence for
localized $d$-electronic states within the metallic phase seen for
$0.2 \leq x \leq 0.5$. This possibility is supported by our LDA
calculations and is consistent with the predictions of the
orbital-selective Mott-transition scenario.

We acknowledge useful discussions with M. Braden, P. Schlottmann,
and P. B. Littlewood. LB acknowledges support from the NHMFL
in-house program and ZF from NSF DMR 0433560. The NHMFL is
supported by NSF-DMR-0084173. Work at Kyoto was supported in part
by Grants-in-Aid for Scientific Research from JSPS and for the
21st Century COE ``Center for Diversity and Universality in
Physics" from MEXT of Japan.


\begin{thebibliography}{}
\bibitem{satoru} S. Nakatsuji, and Y. Maeno, Phys.\ Rev.\ Lett.\ \textbf{84}, 2666 (2000)
\bibitem{satoruprb}S. Nakatsuji, and Y. Maeno, Phys.\ Rev.\ B
\textbf{62}, 6458 (2000).
\bibitem{yoshi} Y.\ Maeno \emph{et al.}, J.\ Phys.\ Soc.\ Jpn.\ \textbf{66}, 1405 (1997).
\bibitem{andy} A.\ P.\ Mackenzie \emph{et al.}, Phys.\ Rev.\ Lett.\ \textbf{76} 3786 (1996).
\bibitem{satorujpsj} S.\ Nakatsuji \emph{et al.}, J.\ Phys.\ Soc.\ Jpn.\ \textbf{66}, 1868 (1997).
\bibitem{bergemann} C.\ Bergemann, \emph{et al.}, and Y. Maeno, Phys. Rev. Lett. \textbf{84}, 2662 (2000) and references therein.
\bibitem{satoru2} S.\ Nakatsuji \emph{et al.}, Phys.\ Rev.\ Lett.\ \textbf{90}, 137202
(2003).
\bibitem{anisimov} V.\ I.\ Anisimov \emph{et al.}, Eur.\ Phys.\ J B \textbf{25}, 191
(2002).
\bibitem{liebsch} A.\ Liebsch, Phys.\ Rev.\ Lett.\ \textbf{91}, 226401
(2003); Europhys. Lett. \textbf{63}, 97 (2003); Phys.\ Rev.\ B
\textbf{70}, 165103 (2004).
\bibitem{koga} A.\ Koga, N.\ Kawakami, T.\ M.\ Rice, and M.\ Sigrist, Phys.\ Rev.\ Lett.\ \textbf{92}, 216402
(2004).
\bibitem{okamoto} S.\ Okamoto and A.\ J.\ Millis, Phys.\ Rev.\ B \textbf{70},
195120 (2004).
\bibitem{arpes} S.\ C.\ Wang, \emph{et al.}, Phys.\ Rev.\ Lett.\ \textbf{93}, 177007
(2004).
\bibitem{wosnitzaishiguro} T.\ Ishiguro, K.\ Yamaji, \emph{Organic Superconductors},(Springer,
Berlin, 1990); J.\ Wosnitza, \emph{Fermi Surfaces of
Low-dimensional Organic Metals and Superconductors}, (Springer,
Berlin, 1996).
\bibitem{ohmichiamro} E.\ Ohmichi, \emph{et al.}, Phys.\ Rev.\ B \textbf{59}, 7263
(1999).
\bibitem{ohmichiinplane} E.\ Ohmichi, \emph{et al.}, Phys.\ Rev.\ B \textbf{61}, 7101
(2000).
\bibitem{bradenhc} M.\ Braden, private communication.
\bibitem{taillefer} S.\ Y.\ Li, \emph{et al.}, Phys.\ Rev.\ Lett.\ \textbf{93}, 056401
(2004).
\bibitem{braden} M.\ Kriener \emph{et al.}, cond-mat/0408015.
\bibitem{sigrist} M.\ Sigrist and M. Troyer, Eur. Phys. J. B \textbf{39},
207 (2004).
\bibitem{andyHall} L.\ M.\ Galvin, R.\ S.\ Perry, A.\ W.\ Tyler, and A.\ P.\ Mackenzie,
Phys.\ Rev.\ B \textbf{63}, 161102(R) (2001).
\bibitem{gang} J.\ F.\ Karpus, \emph{et al.},
Phys.\ Rev.\ Lett.\ \textbf{93}, 167205 (2004); X.\ N.\ Lin,
\emph{et al.}, \emph{ibid} \textbf{95}, 017203 (2005).
\bibitem{nigelprl} N.\ E.\ Hussey \emph{et al.}, Phys.\ Rev.\ Lett.\ \textbf{76},
122 (1996).
\bibitem{yamaji} K.\ Yamaji, J.\ Phys.\ Soc.\ Jpn. \textbf{58}, 1520 (1989).
\bibitem{lebed}A.\ G.\ Lebed, N.\ N.\ Bagmet and M.\
J.\ Naughton, Phys.\ Rev.\ Lett.\ \textbf{93}, 157006 (2004).
\bibitem{fang} Z.\ Fang, N.\ Nagaosa, and K.\ Terakura, Phys.
Rev. B \textbf{69}, 045116 (2004).


\end{thebibliography}
\end{document}